**Sustainability: Scholarly Repository as an Enterprise**
*ASIST Bulletin*, **October/November 2012**

Oya Y. Rieger

*Editors Summary: The expanding need for an open information sharing infrastructure to promote scholarly communication led to the pioneering establishment of arXiv.org, now maintained by the Cornell University Library. To be sustainable, the repository requires careful, long term planning for services, management and funding. The library is developing a sustainability model for arXiv, based on voluntary contributions and the ongoing participation and support of 200 libraries and research laboratories around the world. The sustainability initiative is based on a membership model and builds on arXiv's technical, service, financial and policy infrastructure. Five principles for sustainability drive development, starting with deep integration into the scholarly community. Also key are a clearly defined mandate and governance structure, a stable yet innovative technology platform, systematic creation of content policies and strong business planning strategies. Repositories like arXiv must consider usability and lifecycle alongside values and trends in scholarly communication. To endure, they must also support and enhance their service by securing and managing resources and demonstrating responsible stewardship.*

Over the past two decades, advancements in information and communication technologies have ushered new modes of knowledge creation, dissemination, sharing and enquiry. Open access has emerged as an alternative and viable publishing model and an increasingly vital component of the scholarly communication infrastructure. The vision of an open and robust information infrastructure aims to facilitate the broad dissemination of research outputs of all types – including research data – to allow use, nullification, refinement and reuse. Creating a ubiquitous, comprehensive and linked research data environment requires a seamless network of content, technologies, policies, expertise and practices. It is also critical to view this scholarly organization as an enterprise that needs to be maintained, improved, assessed and promoted overtime.

Open access does not entirely remove fees and access limitations, but it replaces and reconfigures them for the key stakeholders in the scholarly communication endeavor. As we explore a range of issues related to research data curation and management, it is prudent and timely to consider how these services will be maintained and developed in order to flourish over time. What we may consider an exploratory or pilot initiative often ends up transitioning into production, sometimes with insufficient time to think through implications. Therefore as we envision research data support, it is critical that we consider long-term development and management issues upstream as a component of an enduring service infrastructure. Simply put, sustainability is the capacity to endure. For repositories, sustainability entails long-term maintenance of responsibility, which has technical, socioeconomic, policy and business dimensions and encompasses the concept of stewardship, the responsible management of resource use. At the heart of the concept is the ability to secure resources such as technologies, expertise, policies, visions and standards needed to protect and enhance the value of a service based on the needs of the user community technologies. In the pursuing discussion, based on Cornell's experience with arXiv, I highlight the key premises of sustainability.

**arXiv Sustainability Initiative**
Started in August 1991 by Paul Ginsparg, arXiv.org is internationally acknowledged as a pioneering digital archive and open-access distribution service for research articles. The e-print repository, which moved to the Cornell University Library in 2001, has transformed the scholarly communication infrastructure of multiple fields of physics and plays an increasingly prominent role in mathematics, computer science, quantitative biology, quantitative finance and statistics. As of August 2012, arXiv included over 770,000 e-prints. In 2011 there were 76,578 new submissions and close to 50 million downloads from all over the world. arXiv's operating costs for 2012 are projected to be approximately $550,000, including six FTE staff, server maintenance and networking.

Since January 2010, Cornell University Library (CUL) has undertaken an effort to establish a sustainability model for arXiv. As a three-year interim strategy for 2010-2012, CUL initially established a voluntary institutional contribution model and invited pledges from 200 libraries and research laboratories worldwide that represent arXiv's heaviest institutional users. Since scholars worldwide depend on the stable operation and continued development of arXiv, this strategy aimed to reduce arXiv's dependence on a single institution, instead creating a broad-based, community-supported resource. Keeping open access academic resources such as arXiv sustainable entails not only covering the associated operating costs but also continuing to enhance the resources' value based on the needs of the full range of user communities. Cornell's sustainability initiative has striven to assess and strengthen arXiv's technical, service, financial and policy infrastructure. One of the goals of the business planning initiative has been to engage the institutions that benefit from arXiv to assist in defining the future of the service.

The arXiv membership model, which will be launched in 2013, is an outcome of the three-year planning process and was facilitated with a planning grant from the Simons Foundation. The model is founded on a set of operating principles for arXiv and presents a business model for generating revenues. The background information about the planning activities is available at http://arxiv.org/help/support. The business model is composed of four sources of revenue: Cornell's annual funding of $75,000 per year and indirect expenses (represents 37% of direct expenses), $50,000 per year gift from the Simons Foundation, annual fee income from the member institutions and a $300,000 per year challenge grant from the Simons Foundation based on the revenues generated through membership payments. The substantial gift from the Simons Foundation aims to encourage long-term community support by lowering arXiv membership fees, making participation affordable to a broad range of institutions. Based on institutional usage ranking, the annual fees are set in four tiers within the $1,500-3,000 range.

**Sustainability Principles**
arXiv as a sociotechnical system consists of technical systems and standards – activities and practices involved in developing and using the system and the social arrangements and organizations that provide it with a structural framework. The sustainability planning process aims not only to investigate how to diversify revenue models but also to ensure that arXiv strives to meet a set of operational, editorial, financial and governance principles and can set an example of a transparent and reliable community-supported service. Based on Cornell's experience in planning the future of arXiv, the following discussion proposes five sustainability principles, including the consideration of disciplinary cultures, creation of a governance structure, stability of technological components, creation of content policies and adherence to managerial best practices [1]

***Deep Integration into the Scholarly Community and Scholarly Processes***. Disciplinary characteristics, work practices and conventions of academia play an important role in researchers' assessment and appropriation of information communication technologies. Repository deployment cannot be fully understood without comprehending how a specific technology is embedded in its social context. The information and communication technology integration characteristic of disciplinary communities often mirrors underlying differences in epistemic cultures. arXiv is a scholarly communication forum informed and guided by scientists and the scientific cultures being served. Through Paul Ginsparg's leadership, which is rooted in both the academic and information science communities, the service has consistently focused on the disciplinary cultures represented in the digital repository and on community need [2]. arXiv's wide international acceptance attests the importance of relying on user- and evidence-based strategies for informing IT modifications and enhancements, user support services and associated repository policies.

***Clearly Defined Mandate and Governance Structure***. Although best practices in developing technical architectures and associated processes and policies underpin a digital repository, organizational attributes are equally important. The Trustworthy Repositories Audit & Certification: Criteria and Checklist (TRAC) tool (http://trac.edgewall.org/) emphasizes that organizational attributes affect the performance, accountability and sustainability of repositories. The first criteria in the TRAC assessment tool are governance and organizational viability. Similarly, subject repositories must have clearly defined mandates and associated governance structures to reflect a commitment to the long-term stewardship of a service.

The general purpose of governance is to ensure that an organization has the means to envision its future and the management structures and processes in place to ensure that the plan is implemented and sustained. Good governance is participatory, consensus oriented, accountable, transparent, responsive, efficient, equitable and inclusive. However, it also needs to be nimble and flexible – not allowing any gridlocks or excessive groupthink.

The key editorial, governance and economic tenets of arXiv are delineated by a set of principles (https://confluence.cornell.edu/x/xKSTBw). CUL holds the overall administrative and financial responsibility for arXiv's operation and development, with strategic and operational guidance from its Member Advisory Board (MAB) and its Scientific Advisory Board (SAB). CUL manages the moderation of submissions and user support (including the development and implementation of policies and procedures), operates arXiv's technical infrastructure, assumes responsibility for archiving to ensure long-term access, oversees arXiv mirror sites and establishes and maintains collaborations with related initiatives to improve services for the scientific community through interoperability and tool sharing. MAB is elected from arXiv's membership and advises CUL on issues related to repository management and development, standards implementation, interoperability, development priorities, business planning and financial planning. SAB is composed of scientists and researchers in areas covered by arXiv and provides advice and guidance pertaining to the intellectual oversight of arXiv, with particular focus on the policies and operation of arXiv's moderation system.

***Technology Platform Stability and Innovation***. The existing repository ecology is a complex of architectures and features that are optimized to fulfill the specific needs of institutional, subject or archival repositories. The landscape is becoming even more heterogeneous with the addition of scientific social networking sites that profile local scholarly activities and open data initiatives that focus on data curation models. A critical component of a sustainability plan is to consider this rich context and understand how the service fits within the broader framework. In particular, we need to factor in the following three aspects:

- Interoperability arrangements that link a given repository to related systems, services and communities

- Features that support supplementary information objects such as underlying data, auxiliary multimedia content and research methodologies

- Functionality and arrangements that lower barriers to contributing content to multiple complementary repositories.

Among the critical roles of repositories is facilitating the preservation function. Digital preservation (used interchangeably with *archiving*) refers to a range of managed activities to support the long-term maintenance of bitstreams, thereby ensuring that digital objects are usable. However, ensuring enduring access involves more than bitstream preservation. It must provide continued access to digital content through various delivery methods. As we assess the value of subject repositories, it is important to differentiate between bitstream preservation and preserving access. Also, some subject repositories may opt to institute best practices for managing digital content but may not be in a position to assume full preservation responsibilities. It is therefore critical to assess the clearly defined roles and associated procedures that pertain to this responsibility.

There is often an inherent tension between technological stability and innovation. Although dependability and consistency are important service attributes, also essential is keeping pace with evolving user needs through research and development (R&D) projects. Accordingly, assessing resource needs and planning for investments in staffing and equipment are challenging tasks on both the maintenance and R&D fronts. However, given the uncertainties associated with the development and testing of new features and services, an innovation agenda needs to be carefully thought out in order to ensure that operational stability is not undermined.

The sustainability of arXiv also depends on enabling interoperability and creating efficiencies among repositories with related and complementary content to reduce duplication of efforts. For instance, Cornell has collaborated with the NSF Data Conservancy project (http://dataconservancy.org/) to launch a pilot interface that allows arXiv submitters to upload data associated with their articles directly to the Data Conservancy repository. Links to the data are added in the arXiv record automatically. This service is a research project, however, and arXiv and the Data Conservancy make no guarantee about continued availability of datasets uploaded via this mechanism past the end of 2011.

*Systematic Development of Content Policies*. Content curation and stewardship roles have traditionally been shared by libraries and publishers – publishers with a focus on creation and distribution and libraries specializing in discovery and preservation. In digital information environments, repository services such as arXiv blur the distinction between libraries and publishers. An essential criterion in assessing open access, online resources is the availability of clearly defined collection policies and submission guidelines that reflect the content curation and stewardship role of the hosting institution.

Although arXiv is not peer-reviewed, submissions are reviewed by subject-based moderators. Additionally, an endorsement system is in place to ensure that content is relevant to current research in the specified disciplines. It is also critical to have clearly articulated policies about the copyright status of the deposited materials as well as conflict management processes (such as responding to concerns in regard to rejected submissions). arXiv supplements the traditional publication system by providing immediate dissemination and open access to scholarly articles (which often appear later in conventional journals).

arXiv complements, rather than competes with, the commercial and scholarly society journal publishing market. Among the challenges are ensuring the authority and integrity of e-prints and distinguishing between succeeding versions, such as a pre-print paper and its published version in a scholarly journal. It is therefore critical that the repository and preservation community address the versioning of scholarly articles, tracking them from initial submission to pre-print archive to final publication in a formal scholarly journal. Other challenges include linking the burgeoning corpus of institutional repositories with related subject repositories in order to achieve version control as well as creating a critical mass of related materials on particular topics. Cornell's participation in the Open Researcher and Contributor ID (ORCID) author identifiers initiative (http://www.orcid.org) aims to enable better author linking and allow improvements in ownership claiming.

*Reliance on Business Planning Strategies*. Business plans offer an overall view of a given product and its relevant user segments, key stakeholders, communication channels, competencies, resources, networks, collaborations, cost structures and revenue models. The primary purpose of a business planning process is to convey a clear value proposition to justify investment in a service or product by its potential users. The value proposition describes the benefits that a product or service provides. In other words, value propositions respond to this question: Why should an institution purchase your product or service? Since the focus of the value proposition is on the customer, it should be stated from the end-users' perspective. Value propositions may be based on a range of characteristics such as service features, customer support, product customization and economical pricing. The key challenge in creating a value proposition is addressing the needs of all stakeholders. For instance, in the case of arXiv, the stakeholders include scientists, libraries, research centers, societies, publishers and funding agencies. Although they are likely to have common goals, each group attaches value to a specific aspect of arXiv. For instance, from the end-users' perspective, scientists' highest priority for arXiv is likely to be the robustness and reliability of the repository and access features.

Business models also convey financial plans. In a collaborative business model such as the arXiv Membership Model it is critical to clearly define and justify the pricing model and the budget to understand how revenues are being generated and spent. Maintaining, supporting and further developing a repository involve a range of expenses including management, programming, system administration, curation, storage, hardware, facilities (space, furniture, networking, phone), research and training (such as attending meetings and conferences), outreach and promotion, user documentation and administrative support.

**Looking Ahead**
The arXiv case study illustrates the need to manage repositories holistically by taking into consideration a range of lifecycle and usability issues, as well as factoring in changing patterns and values of scholarly communication ecology. Increasing emphasis on open science and the burgeoning data management mandates usher a complex suite of technology, policy and service needs. We must consider the sustainability requirements upstream and remember that the services we are experimenting and creating now have long-term implications. Bowker et al. [3] argue that understanding the new information infrastructures requires an integrative view that goes beyond studying only technical, organizational or social aspects. Such an integrative approach involves comprehending the emerging infrastructure within the context of day-to-day routines, evolving academic practices and business procedures. Viewing scientific repositories as enterprises will ensure that the emerging scholarly communication structures and practices will be effective, efficient and enduring.

**Resources Mentioned in the Article**

---

*Oya Yildirim Rieger is associate university librarian for digital scholarship and preservation services at Cornell University Library. She oversees the library's digitization, online repository, digital preservation, electronic publishing and e-scholarship initiatives with a focus on needs assessment, requirements analysis, business modeling and information policy development. She can be reached at oyr1<at>cornell.edu.*